%% file: paper.tex
\documentclass[11pt,draftclsnofoot,onecolumn]{IEEEtran}

\input{input.tex}

\usepackage{graphicx}
\usepackage{amssymb}
\usepackage{amsfonts}
\usepackage{amsmath}
\begin{document}
\newtheorem{thm}{Theorem}
\newtheorem{lemma}{Lemma}
\newtheorem{rem}{Remark}
\newtheorem{exm}{Example}
\newtheorem{prop}{Proposition}
\newtheorem{defn}{Definition}
\def\proof{\noindent\hspace{0em}{\itshape Proof: }}
\def\endproof{\hspace*{\fill}~\QED\par\endtrivlist\unskip}
\def\bh{{\mathbf{h}}}
\newcommand{\expeq}{\stackrel{.}{=}}
\newcommand{\expg}{\stackrel{.}{\ge}}
\newcommand{\expl}{\stackrel{.}{\le}}

\title{To Code or Not To Code in Multi-Hop Relay Channels}
\author{Rahul~Vaze and Robert W. Heath Jr. \\
The University of Texas at Austin \\
Department of Electrical and Computer Engineering \\
Wireless Networking and Communications Group \\
1 University Station C0803\\
Austin, TX 78712-0240\\
email: vaze@ece.utexas.edu, rheath@ece.utexas.edu
\thanks{This work was funded by DARPA through IT-MANET grant no. W911NF-07-1-0028.}}

\date{}
\maketitle
\noindent
\begin{abstract}

Multi-hop relay channels use multiple relay stages, each with multiple 
relay nodes, to facilitate communication between a source and 
destination. Previously, distributed space-time coding was used to 
maximize diversity gain. Assuming a low-rate feedback link from the
destination to each relay stage and the source, this paper proposes 
end-to-end antenna selection strategies as an alternative to distributed 
space-time coding. One-way (where only
the source has data for destination) and two-way (where the destination 
also has data for the source)
multi-hop relay channels are considered with both the full-duplex and 
half duplex relay nodes. End-to-end antenna selection strategies are 
designed and proven to achieve maximum diversity gain by using a single antenna 
path (using single antenna of the source, each relay stage and the 
destination) with the maximum signal-to-noise ratio at the destination. 
For the half-duplex case, two single antenna paths with the two best 
signal-to-noise ratios in alternate time slots are used to overcome the 
rate loss with half-duplex nodes, with a small diversity gain penalty. 
Finally to answer the question, whether to code (distributed space-time 
code) or not (the proposed end-to-end antenna selection strategy) in a 
multi-hop relay channel, end-to-end antenna selection strategy and 
distributed space-time coding is compared with respect to several 
important performance metrics.

\end{abstract}

\section{Introduction}
Recently, there has been growing interest in designing
maximum diversity gain achieving coding strategies for multi-hop relay
channels \cite{Laneman2003, Laneman2004, Nabar2004,
 Jing2004d, Jing2006a, Belfiore2007, Yang2007a, Sreeram2008,  Peters2007, Bletsas2006, Vaze2008, Yiu2005, Barbarossa2004, Damen2007, Oggier2006k, Oggier2007b }, where a source
uses $N-1$ relay stages to communicate with its destination
and each relay stage is assumed to have one or more relay nodes.
One class of coding strategies proposed to achieve the maximum diversity gain
in a multi-hop relay channels are distributed space-time block codes
(DSTBC). In DSTBCs, coding is done in space, across antennas, and time by
each relay node in a distributed manner.
Maximum diversity gain achieving DSTBCs have been constructed in
\cite{Laneman2003, Laneman2004, Nabar2004, Jing2004d, Jing2006a, Belfiore2007, Yiu2005, Barbarossa2004, Damen2007, Oggier2006k}
for a two-hop relay channel $(N=2)$ and in
\cite{Yang2007a, Sreeram2008, Vaze2008, Oggier2007b} for a general multi-hop relay channel.

Alternative coding strategies to DSTBCs use antenna selection (AS) and
relay selection (RS) to achieve the maximum diversity gain for
two-hop relay channel \cite{Peters2007, Laneman2003, Laneman2004, Bletsas2006,
Zinan2005, Ibrahim2008, Caleb2007}.
With only a single relay node, the AS strategy of
\cite{Peters2007} chooses a single antenna of the source and a single
antenna of the relay node to transmit the signal to the destination,
such that the received signal-to-noise ratio (SNR) is maximized at the
destination which employs minimum mean square error receiver.
For a two-hop relay channel with multiple relay nodes,
the relay selection strategy of \cite{Laneman2003, Laneman2004, Bletsas2006, Zinan2005, Ibrahim2008}
chooses a single relay node from the
set of all possible relay nodes, which maximizes the SNR
at the destination. In \cite{Caleb2007} a subset of relay nodes is selected that maximize the mutual
 information at the destination.


RS strategies are also used for routing in multi-hop networks
\cite{Park2003,Gui2007,Bohacek2008} to leverage the diversity gain.
In \cite{Park2003,Gui2007,Bohacek2008} the route is selected to maximize the
SNR of the worst link and a decode and forward (DF) strategy
is used at each relay node on the selected route.
These routing protocols have been shown to achieve the maximum diversity gain
of multi-hop networks in some special cases.

The primary advantages of AS and RS strategies over DSTBCs are that
they require a minimal number of active antennas and
reduce the encoding and decoding complexity.
Maximum diversity gain achieving AS and RS strategies are only known for
a two-hop relay channel and it is not clear, whether AS and RS
strategies can also achieve maximum diversity gain in a general
multi-hop relay channel. We answer this question in this paper
and propose an end-to-end antenna selection (EEAS) strategy for
a multi-hop relay channel that is shown to achieve the maximum diversity gain.
Thus, we show that distributed space time block coding is not
necessary in a multi-hop relay channel and maximum diversity gain can
be achieved without any space-time coding, when small amount of limited
feedback \cite{Love2005} is available from the destination.
Moreover, we also describe several advantages of using our
proposed EEAS strategy over DSTBC in a
multi-hop relay channel.

In this paper we design EEAS strategies to maximize
diversity gain in the full-duplex multi-hop relay
channel, where each node can transmit and receive at the same time, and the
half-duplex multi-hop relay channel where each node can either transmit or receive at any given
time. We also consider a two-way multi-hop relay channel where the destination
also has data to send to the source.
Throughout this paper we assume that the destination has
channel state information (CSI) for all the channels in the receive mode and
for each EEAS strategy the path selection is done at the destination
using its CSI, and the index of the path to be used is communicated by the destination to the source and each relay stage using a low rate feedback link.
\begin{figure}
\centering
\includegraphics[width= 3.5in]{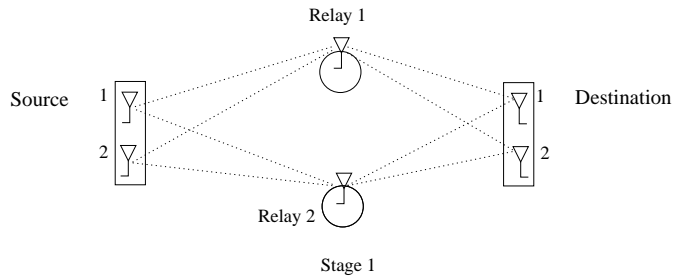}
\caption{A Two-Hop Relay Channel Example}
\label{2hopex}
\end{figure}

We define a single antenna path in a multi-hop
relay channel as a communication channel from the source to the destination,
where only a single antenna of the source, a single antenna of each
relay stage and a single antenna of the destination is used to transmit
the signal from the source to the destination.
We consider amplify and forward (AF) strategy at each relay node,
since the DF strategy limits the multiplexing gain of the multi-hop relay
channel.

For the full-duplex multi-hop relay channel we propose an EEAS strategy,
which chooses a single antenna path over all other single antenna paths, that maximizes the SNR at the
destination. We prove that this EEAS strategy achieves the maximum diversity gain in
a full-duplex multi-hop relay channel by showing that in a multi-hop relay channel,
the maximum number of single antenna paths that do not share any common edges is equal to the upper bound on the
diversity gain of a multi-hop relay channel \cite{Yang2007a}.
Therefore, by selecting the single antenna path that has the maximum SNR at the
destination, maximum diversity gain can be achieved in a full-duplex multi-hop relay channel.
Note that channel gain on each single antenna path is a product of Gaussian scalar channels.
Since the diversity gain of the product of Gaussian scalar channels is $1$ \cite{Sreeram2008}, it follows that the
EEAS strategy achieves the maximum diversity gain.

To gain an intuitive understanding of the result, we provide
the following illustrative example.
Consider the two-hop relay channel as shown in Fig. \ref{2hopex}.
where each single antenna path can be represented
as a $2$-tuple $(e_{ij},\  e_{jk})$,
and $e_{ij}$ is an edge joining the $i^{th}$ source antenna
to the $j^{th}$ relay antenna $i=1,2,\ j=1,2$ and $e_{jk}$ is an
edge joining the $j^{th}$ relay stage antenna and the $k^{th}$ destination
antenna $j=1,2, \ k=1,2$.
Thus, there are $8$ single antenna
paths in total from the source to the destination through $2$ relay nodes
(denoted by dotted lines).
It is worth noting, however, that
there are only $4$ single antenna paths from the source to the destination
that do not have any common edges, i.e.
paths $(11, 11), \ (12, 21), \ (21, 12), \ (22, 22)$. Hence
the channel coefficients on these $4$ paths are independent.
Thus, if the source chooses the path (one out of these $4$ paths)
that maximizes the SNR at the destination,
there is a potential diversity gain of $4$ to be leveraged,
similar to the AS diversity gain in point-to-point multiple antenna channel
\cite{Molisch2004,Sanayei2004}. Furthermore, from \cite{Yang2007a},
it can also be shown that $4$ is also an upper bound on
the diversity gain for the two-hop relay channel Fig. \ref{2hopex}.
Thus, the proposed EEAS strategy for the full-duplex multi-hop
relay channel can achieve maximum diversity gain in two-hop relay channel
Fig. \ref{2hopex}.

Next, we consider a half-duplex multi-hop relay channel, where
each node can only work in half-duplex mode. Clearly, if we
use the EEAS strategy proposed for the
full-duplex case in a half-duplex multi-hop relay channel,
the spectral efficiency is reduced by a factor of $2$. This is because
each antenna on the chosen single antenna path can either transmit or
receive at any given time.
Thus, for a half-duplex multi-hop relay channel, we propose
a different EEAS strategy that alternatively uses two
single antenna paths that have the two best SNRs at the destination, e.g.
the single antenna path with the maximum SNR is used in odd time slots and
the single antenna path with the next best SNR in the even time slots.
We prove that by paying a small price in terms of diversity gain (in comparison
to full-duplex case), this strategy can achieve full-duplex rates
in half-duplex multi-hop relay channel.

Finally, we also consider a two-way multi-hop relay channel where two nodes $T_1$ and $T_2$
want to exchange information with each other via multiple relay stages.
The multiple antenna two-way two-hop relay channel, where
$T_1$, $T_2$, and the relay nodes have multiple antennas, was introduced in
\cite{Rankov2005}. Most of the work on the
multiple antenna two-way two-hop relay channel has been focussed on finding the capacity region \cite{Rankov2005,Rankov2006,Boche2007,Tarokh2007,Kramer2003}
and to the best of our knowledge no work has been reported on the maximizing the diversity gain of two-way relay channel
either for two-hop or multi-hop case.
An example of a two-way relay channel is the downlink and uplink in cellular networks where both the base-station and the
user needs to exchange information with each other.
Let $p$ be the single antenna path from $T_1 \rightarrow T_2$ that has the
maximum product of the norm of the channel coefficient over all paths.
Then, under channel reciprocity assumptions we show that if $T_1$ and $T_2$ use $p$ to transmit their signal to
$T_2$ and $T_1$, respectively, maximum diversity gain can be achieved for both
$T_1\rightarrow T_2$ and $T_2\rightarrow T_1$ communication, simultaneously.
We conclude that the EEAS strategy doubles the
rate of information transfer in a two-way multi-hop relay channel
without any loss in diversity gain.

Our results show that the proposed EEAS strategy 
achieves the maximum diversity gain
in a multi-hop relay channel similar to DSTBCs
 \cite{Yang2007a, Sreeram2008, Vaze2008}. It is not clear, however, which one is the
better strategy among the two. Therefore, a natural question arises as to
whether one should code (DSTBC) or not (proposed EEAS strategy)
in a multi-hop relay channel? To answer this question, we provide a 
comparison of both these
strategies with respect to various important performance metrics.

{\it Organization:} The rest of the paper is organized as follows.
In Section \ref{sec:sys}, we describe the system model for the multi-hop
relay channel and summarize the key assumptions.
We review the diversity multiplexing (DM)-tradeoff for multiple antenna
channels and an upper bound on the
DM-tradeoff of multi-hop relay channel in Section \ref{sec:dmt}.
In Section \ref{sec:full-dup} our EEAS strategy for the full-duplex
multi-hop relay channel is described and shown to achieve the maximum diversity gain.
In Section \ref{sec:halfdup} and \ref{sec:twoway} we discuss antenna
selection strategies for two-way multi-hop relay channel and half-duplex
multi-hop relay channel and analyze their diversity gains. Some numerical results are provided
in Section \ref{sec:sim}.
Final conclusions are made in Section \ref{sec:conc}.

{\it Notation:}
We denote by ${\bA}$ a matrix, ${\bf a}$ a vector and $a_i$ the
$i^{th}$ element of ${\bf a}$.
The determinant and trace of matrix  ${\bf A}$ is denoted by
$\det({\bA})$ and $\tr({\bA})$.
The field of real and complex numbers is denoted by $\bbR$ and $\bbC$,
respectively. The set of natural numbers is denoted by $\bbN$.
The space of $M\times N$ matrices with complex entries is denoted by
${\bbC}^{M\times N}$.
The Euclidean norm of a vector $\bf a$ is denoted by $|\ba|$.
The superscripts $^T, ^{\dag}$ represent the transpose and the transpose
conjugate.
The cardinality of a set ${\cal S}$ is denoted by $|{\cal S}|$.
The expectation of function $f(x)$ with respect to $x$
is denoted by ${\bbE}_{x}(f(x))$.
A circularly symmetric complex Gaussian random variable $x$ with zero mean and
variance $\sigma^2$
is denoted as $x \sim {\cal CN}(0,\sigma)$.
We use the symbol $\expeq$ to represent exponential equality i.e.,
let $f(x)$ be a
function of $x$, then  $f(x) \expeq x^a$ if $\lim_{x\rightarrow \infty}\frac{\log(f(x))}{\log x} = a$ and similarly $\expl$ and $\expg$ denote the exponential
less than or equal to and greater than or equal to relation, respectively.
To define a variable we use the symbol $\bydef$.

\section{System Model}
\label{sec:sys}
\begin{figure}
\centering
\includegraphics[width= 7in]{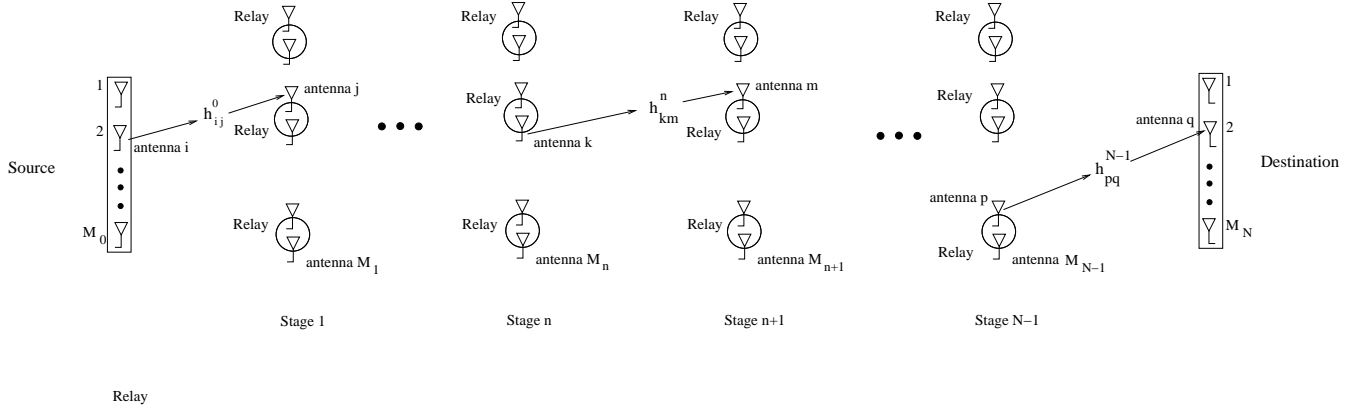}
\caption{System Block Diagram for Multi-Hop Relay Channel}
\label{blkdiag}
\end{figure}

We consider a multi-hop relay channel where a source terminal with $M_0$
antennas wants to communicate with a destination terminal with
$M_N$ antennas via $N-1$ stages of relays as shown in Fig. \ref{blkdiag}.
The $n^{th}$ relay stage has $R_n$ relays and the $j^{th}$ relay
of $n^{th}$ stage has $M_{jn}$ antennas $n=1,2,\ldots,N-1$.
The total number of antennas in the
$n^{th}$ relay stage is $M_n \bydef \sum_{j=1}^{R_n}M_{jn}$.
We assume that the relays do not generate their own data. We assume that
the CSI is only known at the destination and none of the relays have any CSI.
To keep the
relay functionality and relaying strategy simple we do not allow relay nodes to
cooperate among themselves.
We assume that there is no direct path
between the source and the destination. This is a reasonable
assumption for the case when relay stages are used for coverage
improvement and the signal strength on the direct path is very weak.
We also assume that there is no direct path between relay stage $n$
and $n+2$, since otherwise, we can remove the $n+1^{th}$ relay stage
and consider a multi-hop wireless network with $N-2$ relay stages.
We consider both the full-duplex and half-duplex multi-hop relay channel,
where by full-duplex we mean that each node can transmit and receive at the
same time, while in half-duplex case each node can either transmit or receive
at any given time. Unless explicitly stated, we assume the multi-hop
relay channel to be full-duplex.
We also consider a two-way multi-hop relay channel where the destination also
has some data for the source.

As shown in Fig. \ref{blkdiag}, the channel coefficient between the
$i^{th}$ antenna of stage $n$ and $j^{th}$ antenna of stage $n+1$
is denoted by $h^n_{ij}, \ i=1,2,\ldots,M_n, \ j=1,2,\ldots,M_{n+1}\ n=0,1,\ldots,N-1$. The channel coefficient between the $k^{th}$ antenna of stage $n$ and 
the $l^{th}$ antenna of stage $n$ is denoted by $g^n_{kl}, \ k,l=1,2,\ldots,M_{n}, k\ne l, \ n=0,1,\ldots,N-1$.
Stage $0$ represents the source and stage $N$ the destination.

Only for the case of two-way multi-hop relay channel
we assume reciprocity for the channel coefficients, i.e.
the channel between the $j^{th}$ antenna of stage $n+1$ and $i^{th}$
antenna of stage $n$ is $h^n_{ij}$ and the channel between the $l^{th}$
and the $k^{th}$ antenna of relay stage $n$ is $g_{kl}^n$.
This is a reasonable assumption for time-division duplex system,
where calibration is employed at antennas of the adjacent relay stages.

We assume that only the destination knows $h^n_{ij}, g^n_{kl}\ \forall \ i=1,2,\ldots,M_n, \ j=1,2,\ldots,M_{n+1}, k,l=1,2,\ldots,M_{n}, k\ne l, \
n=0,1,\ldots,N-1$. One possible way of acquiring CSI at the destination is by using pilot transmissions from the source
and each relay stage, however, we do not explore the practicalities of
this assumption in this paper. For all the EEAS strategies discussed in this paper, we assume that the destination computes the
end-to-end single antenna path for transmission depending on the relevant metric, and the index of the chosen path is
fed back to the source and each relay stage using a low bit-rate feedback link from the destination.
We assume that $h^n_{ij}, g^n_{kl} \in {\mathbb C}$ are
independent and identically distributed (i.i.d.) ${\cal CN}(0,1)$ entries for all $i,j,n$ to keep the analysis simple
 and tractable. We assume that all these channels are
frequency flat, block fading channels, where the channel coefficients remain
constant in a block of time duration $T_c \ge N$ and change independently from block to block.


\subsection{Problem Formulation}
In this paper we focus on designing EEAS strategies to
achieve the maximum diversity gain in a multi-hop relay channel. Let
${\cal C}$ be a coding strategy for a multi-hop relay channel, then
 the diversity gain $d_{\cal C}$ of ${\cal C}$ is defined
as \cite{Tarokh1999a, Jing2004d}
\[d_{\cal C} = -\lim_{\SNR\rightarrow \infty}\frac{\log{P_e\left(\SNR\right)}}
{\log{\SNR}},
\]
where $P_e\left(\SNR\right)$ is the pairwise error probability and $\SNR$ is the received SNR at the destination
using the coding strategy ${\cal C}$.

To identify the limit on the maximum possible
diversity gain in a multi-hop relay channel, we review the
diversity-multiplexing (DM) tradeoff
\cite{Zheng2003} formulation for multiple antenna channels and an
upper bound on the DM-tradeoff of multi-hop relay channel in the next section.
We will also use the DM-tradeoff formulation to show that the EEAS strategies
proposed in this paper achieve the maximum diversity gain in a multi-hop relay
channel.

\subsection{Review of DM-Tradeoff}
\label{sec:dmt}
Consider a multi-antenna fading channel with $\Nt$ transmit and $\Nr$ receive
antennas, with input output relation
\begin{equation}
\label{dmtrx} \br = \sqrt{P}\bH\bs + \bn,\end{equation} where
$P$ is the power transmitted by the source, $\bs\in
\bbC^{\Nt\times1}$ is the transmitted signal with unit power, $\bH
\in \bbC^{\Nr\times \Nt}$ is the matrix of channel coefficients and
$\bn$ is the circularly symmetric complex Gaussian noise with zero mean and $\sigma^2$
variance. Let $\SNR \bydef \frac{P}{\sigma^2}$.
Then the outage probability $P_{out}(R)$ is defined as
\[P_{out}(R) \bydef P\left(I(\bs;\br)\le R\right),\]
where $R$ is the rate of transmission and
$I(\bs;\br)$ is the mutual information between $\bs$ and $\br$ \cite{Cover2004}.

Following \cite{Zheng2003}, let ${\cal C}(\SNR)$ be a family of codes one for
each $\SNR$. Then we define $r$ as the multiplexing gain of
${\cal C}(\SNR)$ if the data rate $R(\SNR)$ of ${\cal C}(\SNR)$ scales as $r$ with respect to
$\log \SNR$, i.e.
\[\lim_{\SNR\rightarrow \infty}\frac{R(\SNR)}{\log \SNR} =r\]
and $d(r)$ as the rate of fall of probability of error $P_e$ of ${\cal
C}(\SNR)$ with respect to \SNR, i.e.
\[P_{e}(\SNR) \expeq \SNR^{-d(r)}.\]
Clearly, $d(r)$ can be interpreted as diversity gain at rate $R=r\log\SNR$.
Recall that, earlier we defined the diversity gain $d_{\cal C}$
of a coding scheme ${\cal C}$ without changing the rate of transmission
i.e. $r=0$, as
\[d_{\cal C} = - \lim_{\SNR\rightarrow \infty}\frac{\log\left(P_e(\SNR)\right)}{\log \SNR}.\]
Thus, $d_{\cal C} = d(0)$ from the DM-tradeoff formulation.
Let $d_{out}(r)$ be the $\SNR$ exponent of $P_{out}$ with rate of transmission
$R$ scaling as $r\log \SNR$, i.e.
\[P_{out}(r\log \SNR)\expeq \SNR^{-d_{out}(r)},\]
then  using the analysis of \cite{Zheng2003} or the compound channel argument
of \cite{Tavildar2006}, it can be shown that
\[P_{e}(\SNR) \expeq P_{out}(r\log \SNR)\]
for random Gaussian signals and
\[d(r) = d_{out}(r).\]
Thus, to compute the diversity gain of a coding scheme it is sufficient
to compute $d_{out}(r)$. The SNR exponent $d_{out}$ of the outage probability
has also been found in \cite{Zheng2003} for (\ref{dmtrx}), which is given by the piecewise
linear function connecting the points $\left(r,d_{out}(r)\right)$,
$r=0,1,\ldots,\min
\{\Nt,\Nr\}$ where
\begin{equation}
\label{dmtmimo}
d_{out}(r) = (\Nt-r)(\Nr-r).\end{equation}
The maximum value of $r$ for which $d(r)\ge 0$ is called the maximum
multiplexing gain.

Next we present an upper bound
on the DM-tradeoff of the multi-hop relay channel obtained in \cite{Yang2007a}.
\begin{lemma}
\label{upbounddmt}
The DM-tradeoff curve $\left(r,d_{out}(r)\right)$ is upper bounded by
 the piecewise linear function connecting the points
$\left(r^n,d^n_{out}(r)\right)$, $r=0,1,\ldots,\min\{M_n,M_{n+1}\}$ where
\[d^n_{out}(r) = (M_n-r)(M_{n+1}-r).\] for each $n=0,1,2,\ldots,N-1$.
\end{lemma}

The upper bound on the DM-tradeoff of multi-hop relay channel is obtained
by using the cut-set bound \cite{Cover2004} and allowing all relays in
each relay stage to cooperate.
Using the cut-set bound it follows that the mutual information between the
source and the destination cannot be more than the mutual information
between the source and any relay stage or between any two relay stages.
Moreover, by noting the fact that
mutual information between any two relays stages is upper bounded
by the maximum mutual information of a point-to-point MIMO channel
with $M_n$ transmit and $M_{n+1}$ receive antennas $, \ n=0,1,\ldots,N-1$,
the result follows from (\ref{dmtmimo}).

Recall that in our model we do not allow cooperation between relay nodes.
Therefore, designing coding strategies that achieve the optimal DM-tradeoff
and in particular maximum diversity gain in a multi-hop relay channel
is a difficult problem.
The difficulty is two-fold, proposing a ``good" coding strategy and
analyzing its DM-tradeoff. We address both these problems in this paper.
First we introduce a directed multi-hop network to simplify the
DM-tradeoff analysis in a multi-hop relay channel.
Then we propose an EEAS strategy for
multi-hop relay channels and analyze its DM-tradeoff using the directed
multi-hop relay channel.

Next, we define the directed multi-hop relay channel and illustrate
how it simplifies the DM-tradeoff analysis of any coding strategy in a
 multi-hop relay channel.
\begin{defn} A multi-hop relay channel  is called a {\bf directed
multi-hop relay channel}, if any relay of relay stage $n$
can only receive signal from its
preceding stage $n-1$ and not from any relay of stage $n$ or $n+1$.
\end{defn}
\begin{rem}
The DM-tradeoff upper bound
(Theorem \ref{upbounddmt}) also holds for a directed multi-hop relay channel,
since it is obtained by isolating relay stage $n$ and
$n+1, \ n=0,1,\ldots, N-1$ which is a point-to-point MIMO channel and
consequently a directed multi-hop relay channel with number of relay stages equal to zero.
\end{rem}
In the next Lemma we show that the DM-tradeoff analysis of a
coding strategy is simpler with a directed multi-hop relay channel
than with a multi-hop relay channel and the DM-tradeoff of a
coding strategy in a multi-hop relay channel is lower bounded by the
its DM-tradeoff in a directed multi-hop relay channel.
\begin{lemma}
\label{directednetwork}
The DM-tradeoff of coding strategy ${\cal C}$ in a
multi-hop relay channel is lower bounded by its DM-tradeoff
in a directed multi-hop relay channel.
\end{lemma}
\begin{proof} Let $\bx_{t}, \ t=1,\ldots, T, \ T \le T_c$ be
the transmitted signal by source with coding strategy ${\cal C}$ at time $t$.
Then the received signal $r_{N+t}$ at the destination of a multi-hop relay channel at
time $N+t$ is
\begin{eqnarray*}\br_{N+1} &=& \bH_1\bx_1 +\bv_{N+1}\\
\br_{N+2} &=& \bH_1\bx_{2} + \bH_2\bx_{1} + \bv_{N+2} \\
& \vdots & \\
\br_{N+T} &=& \bH_1\bx_{T} + \bH_{2}\bx_{T-1} + \ldots + \bH_T\bx_{1} +  \bv_{N+T},
\end{eqnarray*}
where $\bH_t, \ t=1,\ldots,T$ is the matrix whose entries are functions
of channel coefficients $h_{ij}^n, g_{lm}^n, , \ i=1,\ldots, M_n, \
j=1,\ldots,M_{n+1}, \ l\ne m, \ l,m = 1,\ldots,M_n \ n=1,\ldots,N-1$
and $\bv_{N+t}, \ t=1,2,\ldots, T$ is the additive white complex
Gaussian vector received at time $N+t$.
Combining the received signals,
\[\br \bydef \left[\br_{N+1} \ \ldots \ \br_{N+T}\right]^T =
\underbrace{\left[\begin{array}{cccc} \bH_1 & {\bf 0} & {\bf 0} & {\bf 0}\\
\bH_2 & \bH_1 & {\bf 0} & {\bf 0}\\
\vdots & \vdots &\ddots & \vdots \\
\bH_{T} & \bH_{T-1} & \ldots & \bH_1\\
\end{array}\right]}_{\bH^{ud}}\left[\bx_{1} \ \ldots \ \bx_{T}\right]^T +
\left[\bv_{N+1} \ \ldots \ \bv_{N+T}\right]^T.\]
Note that the matrices $\bH_t, \ t=2,\ldots, T$ are due to the interference constraint of the wireless medium, i.e.
in the receive mode relay stage $n$ receives the signals from both the relay stage $n-1$ and $n+1$. Therefore, the
transmitted signal from relay stage $n$ consists of the new symbols received from relay stage $n-1$ and
already transmitted signal received from relay stage $n+1$. Consequently, the signal
propagates back and forth in the multi-hop relay channel and creates a channel with memory at the destination. Due to
this effective channel with memory at the destination it becomes difficult to analyze the diversity gain of any
coding strategy. To simplify this problem we use the directed multi-hop relay channel. Recall that in a
 directed multi-hop relay channel relay stage $n$ can only receive signals from relay stage $n-1$ and as a result signals do not propagate back and forth. 
Therefore the received
signal $r^d_{N+t}$ at the destination of a directed multi-hop relay
channel at time $N+t$ is
\begin{eqnarray*}
\br^d_{N+1} &=& \bH_1\bx_1 +\bv_{N+1}\\
\br^d_{N+2} &=& \bH_1\bx_{2} +  \bv_{N+2} \\
& \vdots & \\
\br^d_{N+T} &=& \bH_1\bx_{T} +  \bv_{N+T}.
\end{eqnarray*}
Combining the received signals,
\[\br^d \bydef \left[\br^d_{N+1} \ \ldots \ \br^d_{N+T}\right]^T =
\underbrace{\left[\begin{array}{cccc} \bH_1 & {\bf 0} & {\bf 0} & {\bf 0}\\
{\bf 0} & \bH_1 & {\bf 0} & {\bf 0}\\
\vdots & \vdots &\ddots & \vdots \\
{\bf 0} & {\bf 0} & \ldots & \bH_1\\
\end{array}\right]}_{\bH^{d}}\left[\bx_{1} \ \ldots \ \bx_{T}\right]^T +
\left[\bv_{N+1} \ \ldots \ \bv_{N+T}\right]^T.\]
Note that the channel matrix $\bH^d$ with a directed multi-hop relay channel
is a diagonal channel, whereas the channel matrix $\bH^{ud}$ with
a multi-hop relay channel is a lower triangular
matrix.  Therefore, it is easy to see that diversity gain analysis
with a directed multi-hop relay channel is simpler than with a multi-hop relay channel.
Moreover, from Theorem $3.3$  \cite{Sreeram2008},
it follows that the DM-tradeoff of ${\cal C}$ with channel $\bH^{ud}$
is lower bounded by the DM-tradeoff of ${\cal C}$ with channel $\bH^d$. Thus, we
conclude that the DM-tradeoff of ${\cal C}$ in a multi-hop relay channel is
lower bounded by the DM-tradeoff of ${\cal C}$ in a directed multi-hop relay channel.
\end{proof}

As specified before, the main focus of this paper is on designing
coding strategies for multi-hop relay channel that can achieve the maximum
diversity gain. As a corollary of Lemma \ref{directednetwork}, thus,
it follows that it is sufficient to
show that a coding strategy achieves the maximum diversity gain in a directed
multi-hop relay channel to conclude that the coding strategy
achieves the maximum diversity gain in a multi-hop relay channel.
Thus, the directed multi-hop relay channel not only simplifies the
diversity gain analysis of a coding strategy but also gives us a sufficient
condition to test its optimality in terms of achieving the maximum
diversity gain in a multi-hop relay channel.

In the next two sections we propose EEAS strategies for
the full-duplex and the half duplex multi-hop relay channel. We derive lower
 bounds on their diversity gain by computing their diversity gain with
a directed multi-hop relay channel.
For the next two sections, whenever we say a multi-hop relay channel we mean
a directed multi-hop relay channel.

\section{Full-Duplex Multi-Hop Relay Channel}
\label{sec:full-dup} In this section we propose an EEAS strategy for
the full-duplex multi-hop relay channel and show that it achieves
the maximum diversity gain. Before introducing our EEAS strategy and
analyzing its diversity gain, we need the following definitions and
Lemma \ref{maxindepath}.

\begin{defn} Let $e^n_{ij}$ be the edge joining antenna $i$ of stage $n$ to
antenna $j$ of stage $n+1$ then a path in a multi-hop relay channel is
defined as the sequence of edges $\left(e^0_{i_0i_1}, e^1_{i_1i_2}, \ldots,
e^{N-1}_{i_{N-1}i_N}\right) \ i_n\in \{0,1,\ldots, M_n\}, \
n\in\{0,1,\ldots,N\}$.
\end{defn}
\begin{figure}
\centering
\includegraphics[height= 2in]{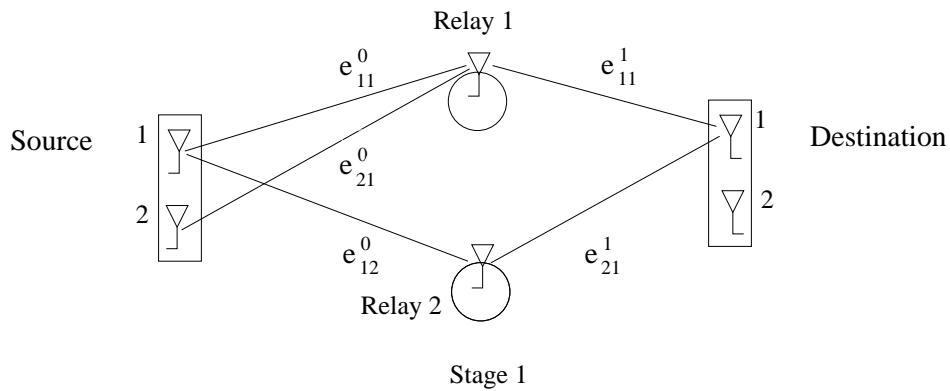}
\caption{An example of a $2$-hop relay channel.}
\label{twohoppaths} 
\end{figure}
\begin{exm}In Fig. \ref{twohoppaths}, $(e^0_{11}, e^1_{11})$ is a path
and so is $(e^0_{21}, e^1_{11})$.
\end{exm}

\begin{defn} Two paths in a multi-hop relay channel are called independent
if they share no common edge.
\end{defn}
\begin{exm} Paths $(e^0_{11}, e^1_{11})$ and $(e^0_{12}, e^1_{21})$ are
independent in Fig. \ref{twohoppaths}.
\end{exm}

In the next lemma we compute the maximum number of independent paths in
a multi-hop relay channel.
\begin{lemma}
\label{maxindepath}
The maximum number of independent paths in a multi-hop relay channel is
\[\alpha \bydef \min\left\{M_nM_{n+1}\right\}, \ n=0,1,\ldots,N-1.\]
\end{lemma}
\begin{proof} (Induction) We use induction on $N$ to prove the result.
For $N=2$, it is easy to verify that the maximum number of independent paths 
is $\min\left\{M_0M_1, M_1M_2\right\}$. Suppose the result is
true for a $k$ hop relay channel, $k\ge 2$. We will prove the Lemma
by showing that the result holds for a $k+1$ hop relay channel also.

Any path $p_{k+1}$ from the source to the
destination of the $k+1$ hop relay channel is of the form $\left(
e^0_{i_0i_1}, e^1_{i_1i_2}, \ldots, e^{k-1}_{i_{k-1}i_{k}},
e^{k}_{i_{k}i_{k+1}}\right), \ i_n\in \{0,1,\ldots, M_n\}$. Since
the number of different edges $e^{n}_{i_{n}i_{n+1}}$ from stage $n$
to stage $n+1, \ n=0,1,\ldots, k$ are $M_nM_{n+1}$, the maximum
number of independent paths in a $k+1$ hop relay channel cannot be
more than $\alpha_{k+1}\bydef \min\left\{M_nM_{n+1}\right\}, \
n=0,1,\ldots,k$. Next, to conclude the proof we show that a set
containing $\alpha_{k+1}$ independent paths exists in a $k+1$ hop
relay channel.

From the induction hypothesis the result is true for the $k$ hop relay channel. Thus, the maximum
 number of independent paths in a $k$ hop relay channel is $\alpha_k \bydef
\min\left\{M_nM_{n+1}\right\}, \ n=0,1,\ldots,k-1$.
Let the set containing these $\alpha_k$ independent paths of a $k$ hop relay channel be $\mathbb{P}_k$, $|\mathbb{P}_k| = \alpha_k$ and any path
which is an element of $\mathbb{P}_k$ be $p_k = \left(e^0_{i_0i_1}, e^1_{i_1i_2}, \ldots, e^{k-1}_{i_{k-1}i_{k}}\right),
\ i_n\in \{0,1,\ldots, M_n\}$.
Let $\alpha_{k} = a_k\lfloor\frac{\alpha_k}{M_k}\rfloor + (M_k-a_k)\lceil\frac{\alpha_k}{M_k}\rceil, \ a_k\in \bbN$.
Then by reassigning the last edge on the paths of set $\mathbb{P}_k$, i.e.
changing the assignment of $e^{k-1}_{i_{k-1}i_{k}}$ in
$p_k \in \mathbb{P}_k$, we can obtain another set of independent paths $\mathbb{P}_k^{r}$ in a $k$ hop relay channel with cardinality
$\alpha_k$, where $\lfloor\frac{\alpha_k}{M_k}\rfloor$ independent paths of $\mathbb{P}_k^{r}$ terminate at
antenna $\ell, \ \ell =1,2,\ldots, a_k$ of the destination of the $k$ hop relay channel and
$\lceil\frac{\alpha_k}{M_k}\rceil$ independent paths of $\mathbb{P}_k^{r}$ terminate at antenna $m, \ m=a_k+1, a_k+2, \ldots,
M_k$ of the destination of the $k$ hop relay channel.
For example, consider Fig. \ref{exindepproof}(a) with a $2$ hop relay channel, where the maximum number of independent paths is $4$ and all of them terminate at antenna $2$ of the destination.
With reassignment, however, it is clear from Fig. \ref{exindepproof}(b)
that we can obtain a set of
$4$ independent paths such that the first and the second antenna of the destination have $1$ path terminating at it while $2$ paths terminate at antenna $3$.
Now we extend the $k$ hop relay channel to $k+1$ hop relay channel,
by distributing the $M_k$ antennas of the destination of $k$ hop
relay channel among $R_k$ relays to form stage $k$ and assuming that
the actual destination is one hop away from stage $k$ with $M_{k+1}$
antennas.
Let $\mathbb{P}^{ir}_k \subset \mathbb{P}^{r}_k$ be the set of independent paths of $\mathbb{P}_k^{r}$ that terminate at antenna $i$ of stage $k$.
Then a set of independent paths $\mathbb{P}^i_{k+1}$ from the source to the
destination of the $k+1$ hop relay channel can be obtained by appending one edge $e^k_{ij}$ (between antenna $i$ of
stage $k$ and antenna $j$ of the destination) to each element of $\mathbb{P}^{ir}_k$
i.e. if $p^{ir}_k \in \mathbb{P}^{ir}_k, \ p^i_{k+1} \in \mathbb{P}^i_{k+1}$ then
$p^i_{k+1} = (p^{ir}_k, e^k_{ij}), \ j\in {1,\ldots, M_{k+1}}$.
Clearly, by the definition of independent paths $|\mathbb{P}^i_{k+1}| = \min\left\{|\mathbb{P}^{ir}_k|, M_{k+1}\right\}$,
 since the number of edges from antenna $i$ of stage $k$ to the
destination of $k+1$ hop relay channel are $M_{k+1}$.
Similarly, we can obtain set of independent paths $\mathbb{P}^i_{k+1}$ for each $i=1,2,\ldots, M_k$.
More importantly, paths belonging to $\mathbb{P}^i_{k+1}$ and $\mathbb{P}^j_{k+1}, \ i\ne j$ are also independent
since $\mathbb{P}_{k}^{ir} \subset \mathbb{P}_k^r, \ i=1,2,\ldots,M_k$ and each path in $\mathbb{P}^r_{k}$ is independent of each other. Thus, we get
a set of independent paths for the $k+1$ hop relay channel $\mathbb{P}_{k+1} = \cup_{i=1}^{M_{k}} \mathbb{P}_{k+1}^{i}$ with cardinality of
$|\mathbb{P}_{k+1}| = \sum_{i=1}^{M_k}|\mathbb{P}^{i}_{k+1}| = \sum_{i=1}^{M_k}\min\left\{|\mathbb{P}^{ir}_k|, M_{k+1}\right\}$.
Since, $|\mathbb{P}^{ir}_k|$ is either $ \lfloor\frac{\alpha_k}{M_k}\rfloor$ or $\lceil\frac{\alpha_k}{M_k}\rceil$,
\[\sum_{i=1}^{M_k}\min\left\{|\mathbb{P}^i_k|, M_{k+1}\right\} = \min\{\alpha_k,\ M_kM_{k+1}\} = \min\{M_nM_{n+1}\}, \ n=0,1,\ldots,k = \alpha_{k+1}.\]
Hence, we have shown that $\alpha_{k+1}$ number of independent paths exits in a $k+1$ hop relay channel.
\end{proof}
\begin{rem}
\begin{figure}
\centering
\includegraphics[height= 1.6in]{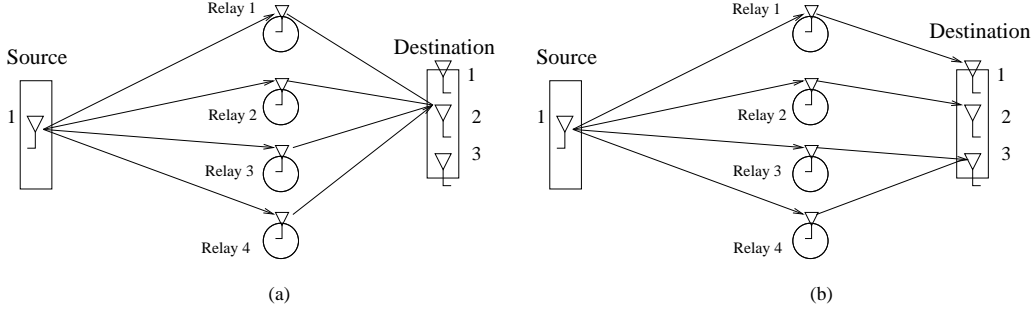}
\caption{An example of a $2$-hop relay channel with reassignment of edges.}
\label{exindepproof} 
\end{figure}
During the preparation of this manuscript we found that Lemma \ref{maxindepath} has been proved independently in \cite{Yang2007a} (Theorem $1$) using a different approach.
\end{rem}

Now we are ready to describe our EEAS strategy for the full-duplex
multi-hop relay channel.
To transmit the signal from source to the destination, a single path in a
multi-hop relay channel is used for communication. How to choose that path is described in the
following. Let the chosen path for the transmission be
$\left(e^0_{i^*_0i^*_1}, e^1_{i^*_1i^*_2}, \ldots,
e^{N-1}_{i^*_{N-1}i^*_N}\right)$, where the signal is transmitted from $i_0^{*th}$
antenna of the source and is relayed through $i_n^{*th}$ antenna of relay stage
$n, \ n=1,2,\ldots N-1$ and decoded by the $i_{N}^{*th}$ antenna
of the destination.
Each antenna on the chosen path uses an AF strategy to
forward the signal to the next relay stage. Thus,
if $x$ is the transmitted signal from the $i_0^{*th}$
source, the received signal at the $i_1^{*th}$ antenna of stage $1$ is
\[r_{i^*_1} = \sqrt{P} h^0_{i^*_0i^*_1}x + v_{i^*_1},\]
where $v_{i^*_1}$ is the complex Gaussian noise with zero mean and unit 
variance, and the antenna $i^*_1$ of stage $1$ transmits
\[t_{i^*_1} = \sqrt{\mu_1}r_{i^*_1}\]
where $\mu_1 =\frac{P}{P+1}$ to ensure that
the average power transmitted is $P$, $\bbE\{|t_{i^*_1}|^2\} = P$.
Similarly, at the $i_n^{*th}$ antenna of stage $n$, the received
signal is scaled by $\mu_n$ (to ensure that
average power transmitted is $P$) and transmitted to antenna
$i^*_{n+1}$ of stage $n+1$.
Thus, with AF by each antenna on the chosen path,
the received signal at the $i_N^{*th}$ antenna of the destination of a 
directed multi-hop relay channel is
\begin{equation}
\label{rxsig}
r_{i^*_N} = \prod_{n=0}^{N-1}\sqrt{\mu_n}\sqrt{P}h^n_{i^*_ni^*_{n+1}}x +
\underbrace{\sum_{m=1}^{N-1}\prod_{k=m}^{N-1}\sqrt{\mu_k}h^k_{i^*_ki^*_{k+1}}v_{i^*_k} + v_{i^*_N}}_{z},
\end{equation}
where $v_{i^*_n}, n=1,2,\ldots, N$ is the complex Gaussian noise with
zero mean and unit variance added at stage $n$ and
$\mu_0=1$.

The EEAS strategy we propose chooses the path that maximizes
the SNR at the destination. Recall that the channel coefficient on the edge
$e^n_{i_q^*i_r^*}$ is $h^n_{i_q^*i_r^*}, \ q =1,2,\ldots, M_n, \ r =1,2,\ldots,M_{n+1},\ n=0,1,\ldots,N-1$.
Let $\sigma^{^*2}$ be the variance of $z$, then,
the EEAS strategy chooses path $(e^0_{i_0^*i_1^*},
e^1_{i_1^*i_2^*}, \ldots, e^{N-1}_{i_{N-1}^*i_{N}^*})$, if
\[\frac{P\prod_{n=0}^{N-1}\mu_n
|h^n_{i^*_ni^*_{n+1}}|^2}
{\sigma^{*2}} = \max_{i_n\in{0,1,\ldots, M_n},
\ n\in\{0,1,\ldots,N-1\}} \frac{P\prod_{n=0}^{N-1}\mu_n
|h^n_{i_ni_{n+1}}|^2}
{\sigma^2}.  \]

Since we assumed that the destination of the multi-hop relay channel has
CSI for all the channels in the receive mode, this optimization can be done at the destination and using
a feedback link, the source and each relay stage can be informed about the index of antennas to use for transmission.

Recall from Lemma \ref{maxindepath} that the maximum number of independent
paths in a multi-hop relay channel is $\alpha$.
It is clear that the path chosen by the EEAS strategy
has better SNR than all the other paths in the multi-hop relay channel.
In particular, the chosen path has better SNR than
atleast $\alpha-1$ independent paths. Thus, by using the proposed antenna
selection strategy, intuitively, one can see that there is a diversity gain of
$\alpha$ to be leveraged. Next, we make this intuition formal, where
we show that the proposed EEAS strategy
achieves the diversity gain of $\alpha$ using the
DM-tradeoff formulation.

\begin{thm}
\label{thm:fulldupdiv}
The proposed EEAS strategy achieves the
maximum diversity gain in a full-duplex multi-hop
relay channel equal to \[\min_{n=0,1,\ldots,N-1}\{M_nM_{n+1}\}.\]
\end{thm}
\begin{proof}
With the proposed EEAS strategy the outage probability of
(\ref{rxsig}) can be
 written as
\[P_{out}(r\log \SNR)= P\left(\log\left(1+\max_{i_n\in{0,1,\ldots, M_n},
\ n\in\{0,1,\ldots,N-1\}} \frac{P\prod_{n=0}^{N-1}\mu_n
|h^n_{i_ni_{n+1}}|^2}
{\sigma^2}\right) \le r\log \SNR\right).\]
Let $\SNR \bydef \frac{P\prod_{n=0}^{N-1}\mu_n}{\sigma^2}$. Clearly,
\begin{eqnarray*}
P_{out}(r\log \SNR) &\le &
P\left(\log\left(1+\max_{\mathbb{P}_N} \SNR\prod_{n=0}^{N-1}
|h^n_{i_ni_{n+1}}|^2\right)
\le
r\log\SNR\right),
\end{eqnarray*}
where $\mathbb{P}_N$ is the set containing maximum number of independent paths in a
multi-hop relay channel.
From Lemma \ref{maxindepath}, $|\mathbb{P}_N|$ is
$\alpha$ and by definition of independent paths,
 channel coefficients on independent paths are independent.
Thus,
\begin{eqnarray*}
P_{out}(r\log\SNR) &= & \prod_{j=1}^{\alpha}P\left(\SNR\prod_{n=0}^{N-1}
|h^n_{i_ni_{n+1}}|^2
\le
\SNR^{-(1-r)}\right). \\
\end{eqnarray*}
$P\left(\prod_{n=0}^{N-1}|h^n_{i_ni_{n+1}}|^2 \le
\SNR^{-(1-r)}\right)$
can be bounded using \cite{Zheng2003},
(formally derived in \cite{Sreeram2008}), and is given by
\[P\left(\prod_{n=0}^{N-1}|h^n_{i_ni_{n+1}}|^2 \le
\SNR^{-(1-r)}\right) \expeq
\SNR^{(1-r)}, \ r\le 1.\]
Thus
\[P_{out}(r\log\SNR) \expl  \SNR^{-\alpha(1-r)}, \ r\le 1\] and
\[d_{out}(r) = \alpha(1-r), r \le 1.\]
From Section \ref{sec:dmt}, it follows that for EEAS strategy
$P_e \expeq \SNR^{-d(r)}$, with $d(r) =d_{out}(r) = \alpha(1-r), \ r\le 1$.
Since $d(r) = \alpha(1-r), \ r \le 1$, the maximum diversity
gain of EEAS strategy (obtained at $r=0$) in a directed multi-hop relay 
channel is
$\alpha = \min_{n=0,1,\ldots,N-1}\{M_nM_{n+1}\}$ which equals the upper bound
on diversity gain from Lemma \ref{upbounddmt}. Hence, using Lemma 
\ref{directednetwork} we conclude that the proposed EEAS strategy achieves the
 maximum diversity gain in a multi-hop relay channel.
\end{proof}

\begin{thm}
The proposed EEAS strategy with AF at each relay achieves the
maximum multiplexing gain in multi-hop relay channel if the source,
or any of the relay stages, or the destination has only a single antenna.
\end{thm}
\begin{proof} Since $d(r) = d_{out}(r) = \alpha(1-r), r \le 1$, the maximum
multiplexing achievable multiplexing gain is $1$ which equals the upper bound
on multiplexing gain from Lemma \ref{upbounddmt}.
\end{proof}

The following remarks are in order.
\begin{rem} Receiver Combining: Note that in the proposed EEAS strategy only 
a single antenna of the destination is used for receiving the signal. 
This simplification is done to keep the diversity gain analysis 
simple. In practice, however, all the antennas of the destination should 
be used for reception to obtain extra array gain.
\end{rem}
\begin{rem} CSI Requirement: Recall that we assumed that the destination
has the CSI for all the channels in a multi-hop relay channel. In light of
lemma \ref{maxindepath} and Theorem \ref{thm:fulldupdiv}, however,
it is clear that the maximum diversity can be achieved even if the destination
has CSI only for the set of $\min_{n=0,1,\ldots,N-1}\{M_nM_{n+1}\}$
independent paths and the path with the best SNR among these independent paths
is chosen for transmission.
Thus, the CSI overhead is moderate for the proposed EEAS strategy.
\end{rem}

\begin{rem} Feedback Overhead: The total number of single antenna paths from
the source to the destination in a multi-hop relay channel are $\prod_{n=0}^{N}
M_n$. Thus, in general, to feedback the index of the best single antenna path
$\log_2\prod_{n=0}^{N} M_n$ bits are needed,
however, since the number of independent paths is only
$\min_{n=0,1,\ldots,N-1}\{M_nM_{n+1}\}$, $\log_2\min_{n=0,1,\ldots,N-1}\{M_nM_{n+1}\}$ bits of feedback is sufficient to achieve the maximum diversity gain.
Therefore the feedback overhead with the proposed EEAS strategy is quite small
and can be realized with a very low rate feedback link.
\end{rem}

{\it Discussion:}
In this section we proposed an EEAS strategy for
a full-duplex multi-hop relay channel.
We analyzed the DM-tradeoff of the proposed EEAS strategy in a directed
multi-hop relay channel and using Lemma \ref{directednetwork} showed that
the proposed EEAS strategy achieves maximum diversity gain in a multi-hop
relay channel.

The result can be interpreted as follows. From Lemma \ref{maxindepath}, we know that there are $\alpha$
independent paths from the source to the destination in multi-hop relay channel.
Thus, if the source chooses the best path out of $\alpha$ independent paths
to communicate with its destination, diversity gain of $\alpha$ can be obtained.
Moreover, from Lemma \ref{upbounddmt}, $\alpha$ is also an upper bound
on the diversity gain in a multi-hop relay channel and thus the proposed EEAS
strategy is optimal in the sense of achieving the maximum diversity gain.

The proposed EEAS strategy is shown to achieve maximum
multiplexing gain of $1$. This is intuitive,
since only one stream is transmitted from the source to the
destination.
Note that for the case of single antenna source or single antenna
destination or any relay stage with a single antenna, the proposed
EEAS strategy is optimal is terms of maximizing the multiplexing gain also.

Recall that DSTBCs constructed in \cite{Vaze2008, Sreeram2008, Yang2007a}
also achieve maximum diversity gain in a multi-hop relay channel.
There are several advantages of using EEAS over DSTBCs, however, such as
reduced noise at the destination, less total power used,  minimal
number of active antennas, minimum latency and minimal decoding complexity,
to name a few. A detailed comparison is provided in the Conclusions section.

\section{Half-Duplex Multi-Hop Relay Channel}
\label{sec:halfdup}
In the previous section we assumed that all nodes in the multi-hop relay
channel are full-duplex, however, full-duplex nodes are difficult
to realize in practice. To address this practical limitation, in this section
we consider a multi-hop relay channel where each node can only work in
half-duplex mode.

It is easy to see that by using the EEAS strategy proposed
for the full-duplex case in a half-duplex multi-hop relay channel,
the DM-Tradeoff curve is given by
$d(r)= \alpha(1-2r)$, since half the time the source and the destination are
silent. Thus, there is a spectral efficiency loss by a
factor of $1/2$ by using the EEAS strategy proposed
for the full-duplex case in a half-duplex multi-hop relay channel.

To improve the rate of transmission with half-duplex nodes,
 we propose an EEAS strategy that uses two paths that have the two best SNRs
at the destination, in alternate time slots,
e.g. the path with the maximum SNR is
used in odd time slots and the path with the next best SNR in
the even time slots. We show that by paying a small price in
terms of diversity gain (in comparison to full-duplex case), this
strategy can achieve full-duplex rates in half-duplex multi-hop
relay channel.

The two paths $p_1$ and $p_2$ for the half-duplex multi-hop relay channel are
selected as follows. The path
$p_1 = \left(e^0_{i^*_0i^*_1}, e^1_{i^*_1i^*_2}, \ldots,
e^{N-1}_{i^*_{N-1}i^*_N}\right)$ is the first chosen path, if
\[\frac{P\prod_{n=0}^{N-1}\mu_n
|h^n_{i^*_ni^*_{n+1}}|^2}
{\sigma^{*2}} = \max_{i_n\in{0,1,\ldots, M_n},
\ n\in\{0,1,\ldots,N-1\}} \frac{P\prod_{n=0}^{N-1}\mu_n
|h^n_{i_ni_{n+1}}|^2}
{\sigma^2}\]
and the path $p_2 = \left(e^0_{j^*_0j^*_1}, e^1_{j^*_1j^*_2}, \ldots,
e^{N-1}_{j^*_{N-1}j^*_N}\right)$ is the second chosen path if
\[\frac{P\prod_{n=0}^{N-1}\mu_n
|h^n_{j^*_nj^*_{n+1}}|^2}
{\sigma^{*2}} = \max_{j_n \ne i_n*, \ j_n \in{0,1,\ldots, M_n},
\ n\in\{0,1,\ldots,N-1\}} \frac{P\prod_{n=0}^{N-1}\mu_n
|h^n_{j_nj_{n+1}}|^2}
{\sigma^2}.\]
Thus, $p_1$ has the best SNR among all paths in a multi-hop
relay channel and $p_2$ has the best SNR among all paths
excluding path $p_1$.
\begin{figure}
\centering
\includegraphics[height= 1.6in]{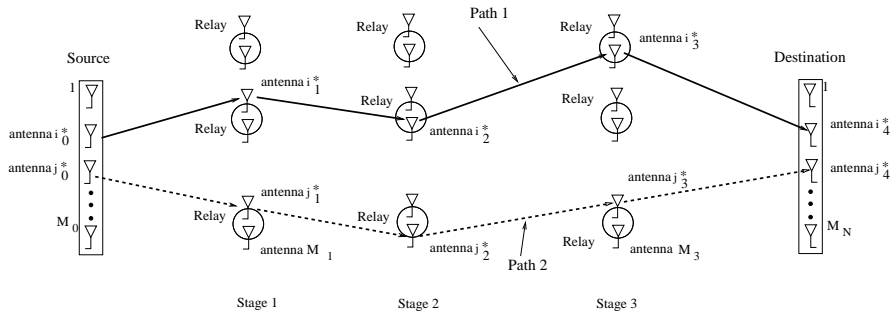}
\caption{An example of EEAS for Half-Duplex $4$-hop Relay Channel}
\label{blkdiaghalfdup}
\end{figure}

From Lemma \ref{maxindepath}, the number of independent paths in a
multi-hop relay channel is $\alpha$. Moreover, the number of
independent paths from the source to the destination that does not
include any antenna that lies on $p_1$ can be calculated by removing
one antenna from each stage and applying Lemma \ref{maxindepath} for
a multi-hop relay channel with $M_n-1$ antennas at each stage. Thus,
the maximum number of independent paths in a multi-hop network that does not
include any antenna that lies on $p_1$ is
\[\beta\bydef \min\{(M_{n}-1)(M_{n+1}-1)\}, \ n=0,1,\ldots,N-1.\]
Then, clearly, path $p_1$ has better SNR than $\alpha-1$ independent paths
and path $p_2$ has better SNR than $\beta-1$ independent paths.

To allow the source and the destination to transmit and receive
continuously in a half-duplex multi-hop relay channel, two paths
$p_1$ and $p_2$ are used alternatively, as follows. In the first
time slot, the source uses antenna $i^*_0$ that lies on path $p_1$
to transmit the signal to the first relay stage which uses antenna
$i^*_1$ to receive the signal. In the next time slot, the source
uses antenna $j^*_0$, that lies on path $p_2$, to transmit the
signal to stage $1$ which uses antenna $j^*_1$ to receive the
signal. In the second time slot, stage $1$ also transmits the signal
it received in the first time slot using antenna $i^*_1$ to stage
$2$ with AF. Similarly, in any given time slot, stage $n$
simultaneously transmits or receives the signal using antenna
$i_n^*$ (path $p_1$) or antenna $j_n^*$ (path $p_2$). In time slot
$t\ge 3$ each node in a half-duplex multi-hop relay channel repeats
the operation it performed in time slot $t-2$.
An illustration of this EEAS strategy is provided in
 Fig. \ref{blkdiaghalfdup} for a $4$-hop relay channel.

To simplify the diversity gain analysis of the proposed EEAS strategy,
we use the directed multi-hop relay channel similar to the
Section \ref{sec:full-dup}. With the proposed EEAS strategy with AF at each
antenna of path $p_1$ or $p_2$, if the source transmits
$x_t$ at time $t$, the received signal at the destination of a directed
multi-hop relay channel in time slots
$N+2t-1$, is
\begin{eqnarray}
\label{halfduprxodd}
r_{N+2t-1} &=&
\prod_{n=0}^{N-1}\sqrt{\mu_n}\sqrt{P}h^n_{i^*_ni^*_{n+1}}x_t +
\underbrace{\sum_{m=1}^{N-1}\prod_{k=m}^{N-1}\sqrt{\mu_k}h^k_{i^*_ki^*_{k+1}}v_{i^*_k} + v_{i^*_N}}_{z_{N+2t-1}},
\end{eqnarray}
and at time $N+2t$ is
\begin{eqnarray}
\label{halfduprxeven}
r_{N+2t} &=&
\prod_{n=0}^{N-1}\sqrt{\mu_n}\sqrt{P}h^n_{j^*_nj^*_{n+1}}x_t +
\underbrace{\sum_{m=1}^{N-1}\prod_{k=m}^{N-1}\sqrt{\mu_k}h^k_{j^*_kj^*_{k+1}}v_{j^*_k} + v_{j^*_N}}_{z_{N+2t}},
\end{eqnarray}
where $t=1,\ 2, \ \ldots$.
Let the variance of $z_{N+2t-1}$ and $z_{N+2t}$ be $\sigma^2$ and
$\SNR \bydef \frac{P\prod_{n=0}^{N-1}\mu_n}{\sigma^2}$, then the
outage probability of (\ref{halfduprxodd}) is
\begin{eqnarray*}
P^{N+2t-1}_{out}(r\log\SNR) & =& P\left(\log\left(1+\prod_{n=0}^{N-1}\SNR |h^n_{i^*_ni^*_{n+1}}|^2\right) \le r\log\SNR\right)\\
&\expeq& P\left(\max_{i_n\in{0,1,\ldots, M_n},
\ n\in\{0,1,\ldots,N-1\}}\prod_{n=0}^{N-1}\mu_n
|h^n_{i_ni_{n+1}}|^2 \le \SNR^{-(1-r)}\right),\end{eqnarray*}
and the outage probability of (\ref{halfduprxeven}) is
\begin{eqnarray*}
P^{N+2t}_{out}(r\log\SNR) & =&  P\left(\log\left(1+\prod_{n=0}^{N-1}\SNR |h^n_{j^*_nj^*_{n+1}}|^2\right) \le r\log\SNR\right)\\
&\expeq& P\left(\max_{j_n\ne i_n^{*}, \ j_n\in{0,1,\ldots, M_n},
\ n\in\{0,1,\ldots,N-1\}} \prod_{n=0}^{N-1}
|h^n_{j_nj_{n+1}}|^2 \le \SNR^{-(1-r)}\right).\end{eqnarray*}
These outage probabilities can be computed by using a similar analysis
 as derived in Section \ref{sec:full-dup} and are given by
\begin{eqnarray*}
P^{N+2t-1}_{out}(r\log\SNR) & \expeq&  \SNR^{-\alpha(1-2r)}, \\
P^{N+2}_{out}(r\log\SNR) & \expeq&  \SNR^{-\beta(1-2r)}.
\end{eqnarray*}
Thus, the effective outage probability for the half-duplex multi-hop relay
channel is
\begin{eqnarray*}
P_{out}(r\log\SNR) & \expeq&  \SNR^{-\beta(1-r)}
\end{eqnarray*}
since $\beta < \alpha$ and dominates the probability of error and
$r$ is in place of $2r$ since the destination receives data in both the
odd and the even time slots. Thus, the DM-tradeoff is
\[d(r) = \beta(1-r),\] and the maximum diversity gain $d(0) = \beta$.
From Lemma \ref{directednetwork}, it follows that
diversity gain of at least $\beta$ is achievable with the proposed EEAS strategy in
 a multi-hop relay channel.

{\it Discussion:} It is clear that our EEAS strategy for
the half-duplex multi-hop relay channel does not achieve the
maximum diversity gain (Lemma \ref{upbounddmt}), since $\beta < \alpha$.
It removes, however, the spectral efficiency loss due to the
half-duplex assumption on the relay nodes.
Thus, with a minimal penalty $\alpha-\beta$ in the diversity gain,
 the proposed EEAS strategy improves the spectral efficiency by a factor of $2$.

%
From the diversity gain analysis, it is clear that with the proposed
EEAS strategy two streams can be sent from the source to the destination in the
alternate time slots which achieve different diversity gains.
This is similar to the idea of diversity embedded space-time codes
\cite{Diggavi2008}, where two different streams are transmitted from a
single source to a single destination with different data rate and
diversity gain requirements.
Thus, our EEAS strategy provides an alternate and simple solution for the
problem considered in \cite{Diggavi2008} applied to a multi-hop relay channel.

Recall that by using the EEAS strategy proposed
for the full-duplex case in a half-duplex multi-hop relay channel, maximum
diversity gain $\alpha$ can be achieved with a spectral efficiency loss.
Thus, there exists a tradeoff between the EEAS strategy proposed in this section
 and the one proposed for full-duplex case.
One offers higher reliability than the other but with a penalty in
spectral efficiency. Therefore a hybrid EEAS strategy can be used
which uses either of these EEAS strategies depending on the
application requirement. For example, for applications such as voice
communication which requires low latency but can tolerate lower
reliability, the EEAS strategy proposed in this section should be
used. Alternatively for applications such as e-mail, EEAS
strategy  proposed for full-duplex case should be used since they
require high reliability but have no latency requirements.



\section{Two-Way Multi-Hop Relay Channel}
\label{sec:twoway}
In both the previous sections we considered one way communication on a
multi-hop relay channel, where a source
wanted to communicate with its destination.
Most wireless networks, however, are two-way in nature, i.e.
the destination also has some data for the source, e.g. downlink and uplink
in cellular wireless networks.
In this section, we consider a two-way
multi-hop relay channel, where two nodes $T_1$ and $T_2$ want to
exchange information with each other using multiple relay stages. The
system model is similar to the one shown in Fig. \ref{blkdiag} with source being
$T_1$ and the destination being $T_2$.

We propose an EEAS strategy for two-way multi-hop relay channels
that achieves maximum diversity gain for both the $T_1 \rightarrow T_2$
and $T_2 \rightarrow T_1$ communication, simultaneously.
We only consider the full-duplex case, the
half-duplex follows along the same lines as half-duplex one way multi-hop
 relay channel case.
In this section we consider a multi-hop relay channel and not the
directed multi-hop relay channel, since otherwise, $T_2's$ message
cannot reach $T_1$.

\begin{figure}
\centering
\includegraphics[height=1.6in]{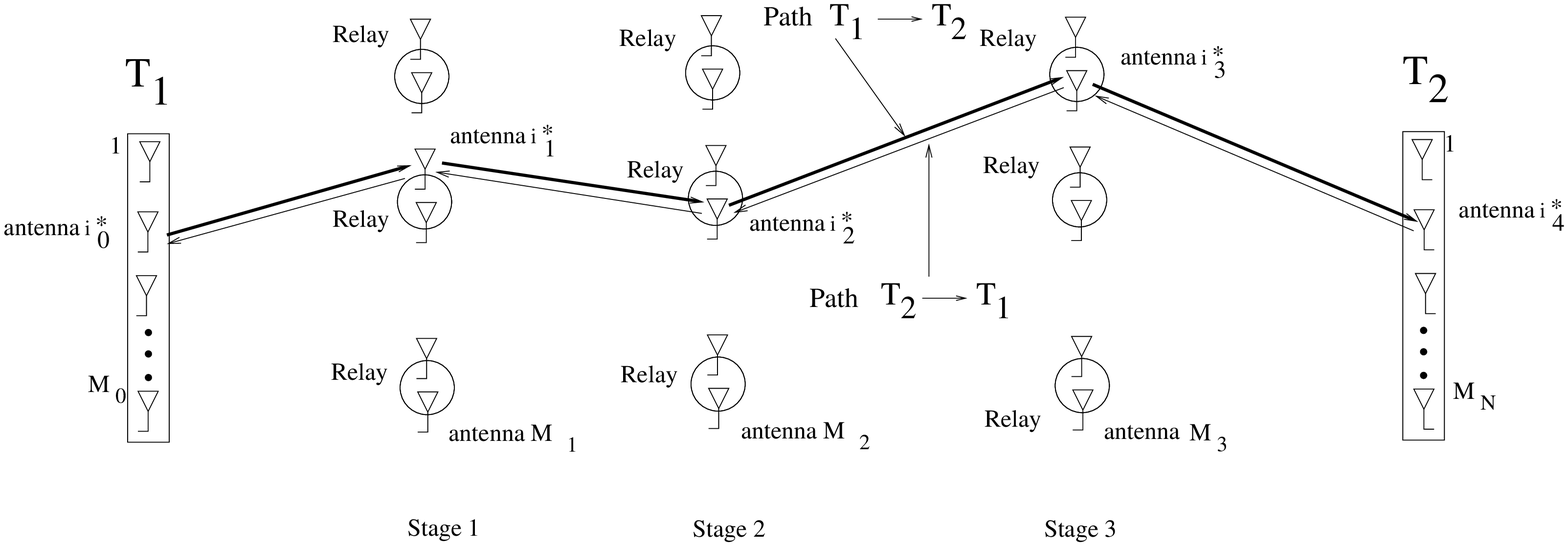}
\caption{An example of EEAS for Two-Way $4$-hop Relay Channel}
\label{blkdiagtwoway}
\end{figure}

To allow both $T_1$ and $T_2$ to send information to each other simultaneously,
we propose an
EEAS strategy that uses a single path from $T_1$ to $T_2$ for
both the $T_1\rightarrow T_2$ and $T_2\rightarrow T_1$ communication.
How to choose the
path is described in the following. Each antenna of stage $n$
that lies on the chosen path uses AF to forward the signal to the antenna of
stage $n-1$ and $n+1$.
Let $p_1$ be the path chosen by the EEAS strategy, where
$p_1 = \left(e^0_{i^*_0i^*_1}, e^1_{i^*_1i^*_2}, \ldots, e^{N-1}_{i^*_{N-1}i^*_N}
\right)$. Then in each time slot, $T_1$ and $T_2$
transmit and receive data using their $i_0^*$ and $i_N^*$ antenna,
respectively. The $i_n^*$ antenna of stage $n$, uses AF to transmit the
signal to the $i_{n-1}^*$ and $i_{n+1}^*$ antenna of stage $n-1$ and $n+1$.
Thus, if the signal transmitted by antenna $i^*_{n}$ of stage $n, \ n=0,1,\ldots,N-1$ at time $t$ be $c_{i^*_{n},t}$,
then the received signal at antenna $i^*_{n}$ of stage $n$ at time $t+1$ is
$r^n_{i^*_{n}t+1} = h^{n-1}_{i^*_{n-1}i^*_n}c_{i^*_{n-1},t} + h^{n+1}_{i^*_{n+1}i^*_n}c_{i^*_{n+1},t} + v^n_{i_n^*}$. The antenna $i^*_{n}$ of stage $n$ then
 transmits $\sqrt{\mu_n}r^n_{i^*_{n}t+1}$, where $\theta_n$ ensures that the average
transmitted power is $P$.
An illustration of this EEAS strategy is provided in Fig. \ref{blkdiagtwoway} for a $4$-hop two-way
relay channel.

Let $x^1_{t}$ and $x^2_{t}$ be the signal transmitted from $T_1$ and $T_2$,
respectively, at time $t$. Then using the EEAS strategy with AF at each chosen
antenna, under the channel reciprocity assumption,
the received signal $y^1_{t+N}$ at $T_1$ and $y^2_{t+N}$ at $T_2$ at time $t+N$ is \begin{eqnarray}
\label{twowayrx}
\nonumber
y^1_{t+N} &=& \sum_{j=0}^{t-1}
\sqrt{\gamma_jP}f^2_j\left(h^n_{i^*_ni^*_{n+1}}\right)x^2_{t-j} +
\sum_{j=0}^{t-1}
\sqrt{\gamma_jP}f^1_j\left(h^n_{i^*_ni^*_{n+1}}\right)x^1_{t-j} \\
&& + \underbrace{\sum_{m=1}^{N-1}\prod_{k=m}^{N-1}\sqrt{\gamma_k}
f^3\left(h^k_{i^*_ki^*_{k+1}}\right)v_{i^*_k} + v_{i^*_N}}_{z^1_{t+N}}
\end{eqnarray}
\begin{eqnarray}
\label{twowayrx1}
\nonumber
y^2_{t+N} &=& \sum_{j=0}^{t-1}
\sqrt{\gamma_jP}f^1_j\left(h^n_{i^*_ni^*_{n+1}}\right)x^1_{t-j} +
\sum_{j=0}^{t-1}
\sqrt{\gamma_jP}f^2_j\left(h^n_{i^*_ni^*_{n+1}}\right)x^2_{t-j} \\
&& + \underbrace{\sum_{m=1}^{N-1}\prod_{k=m}^{N-1}\sqrt{\gamma_k}
f^3\left(h^k_{i^*_ki^*_{k+1}}\right)v_{i^*_k} + v_{i^*_N}}_{z^2_{t+N}},
\end{eqnarray}
where $v_{i^*_k}$ is complex Gaussian noise with
zero mean and unit variance $\forall k$ added by antenna $i^*_k$, $f^m_j\left(h^n_{i^*_ni^*_{n+1}}\right)$ is a function of channel coefficients
$h^n_{i^*_ni^*_{n+1}}$ for $m=1,2,3$, and
$\gamma_i$ is a function of $\mu_n, n=0,1,\ldots,N-1$.

Note that $x_t^1$ and $x_t^2, \forall \ t$ is known at $T_1$ and $T_2$,
respectively, and hence their contribution can be removed from the signals
received at $T_1$ and $T_2$ under perfect CSI assumption. Note that 
$f^m_0\left(h^n_{i^*_ni^*_{n+1}}\right) = \prod_{n=0}^{N-1}h^n_{i^*_ni^*_{n+1}}, \ m=1,2$ and $\gamma_0 = \prod_{n=0}^{N-1}{\mu_n}$.
Removing the contribution of $x_t^1$ and $x_t^2, \forall \ t$ from the received
 signal at $T_1$ and $T_2$ and writing the received signal in time $t=1$ to
$t=T, \ T \le T_c$ in the matrix form,
\begin{eqnarray}
\label{twowayrx}
\left[\begin{array}{c}
y^1_{1+N} \\ y^1_{2+N} \\ \vdots \\y^1_{T+N}
\end{array}\right]
 & = & \underbrace{\left[\begin{array}{cccccc}
\lambda_0 & 0 & 0 & 0 & 0 & 0\\
\lambda^2_1  & \lambda_0 & 0 &  0  &0  & 0\\
\lambda^2_2 &\lambda^2_1  & \lambda_0 & 0 &  0 & 0\\
\lambda^2_3& \lambda^2_2 & \lambda^2_1  & \lambda_0 & 0 &  0 \\
\vdots & \ddots &\ddots & \ddots & \ddots & \vdots \\
 \lambda^2_{T} &\ldots& \lambda^2_3   & \lambda^2_2 & \lambda^2_1 & \lambda_0
\end{array}\right]}_{\bH}
\left[\begin{array}{c}
x^2_{1} \\ x^2_{2} \\ \vdots \\x^2_{T}
\end{array}\right]  + \left[\begin{array}{c}
z^1_{1+N} \\ z^1_{2+N} \\ \vdots \\z^1_{T+N}
\end{array}\right]
\end{eqnarray}
and
\begin{eqnarray}
\label{twowayrx}
\left[\begin{array}{c}
y^2_{1+N} \\ y^2_{2+N} \\ \vdots \\y^2_{T+N}
\end{array}\right]
 & = & \underbrace{\left[\begin{array}{cccccc}
\lambda_0 & 0 & 0 & 0 & 0 & 0\\
\lambda^1_1  & \lambda_0 & 0 &  0  &0  & 0\\
\lambda^1_2 & \lambda^1_1  & \lambda_0 & 0 &  0 & 0\\
\lambda^1_3& \lambda^1_2 & \lambda^1_1  & \lambda_0 & 0 &  0 \\
\vdots & \ddots &\ddots & \ddots & \ddots & \vdots \\
\lambda^1_T & \ldots & \lambda^1_3 & \lambda^1_2 & \lambda^1_1& \lambda_0
\end{array}\right]}_{\bH}
\left[\begin{array}{c}
x^1_{1} \\ x^1_{2} \\ \vdots \\x^1_{T}
\end{array}\right]  + \left[\begin{array}{c}
z^2_{1+N} \\ z^2_{2+N} \\ \vdots \\z^2_{T+N}
\end{array}\right],
\end{eqnarray}
where $\lambda_0 = \prod_{n=0}^{N-1}\sqrt{\mu_nP}h^n_{i^*_ni^*_{n+1}}$,
 $\lambda^m_j = \sqrt{\gamma_jP}f_j^m\left(h^n_{i^*_ni^*_{n+1}}\right), j=1,2,\ldots,T, \ m=1,2$.

Now we describe the EEAS strategy.
The EEAS strategy chooses path $p_1 = (e^0_{i_0^*i_1^*},
e^1_{i_1^*i_2^*}, \ldots, e^{N-1}_{i_{N-1}^*i_{N}^*})$, if
\[P\prod_{n=0}^{N-1}\mu_n
|h^n_{i^*_ni^*_{n+1}}|^2 = \max_{i_n\in{0,1,\ldots, M_n},
\ n\in\{0,1,\ldots,N-1\}} P\prod_{n=0}^{N-1}\mu_n
|h^n_{i_ni_{n+1}}|^2. \] Thus, the EEAS strategy chooses that
path which has the best product of the norm of the channel coefficients
 among all possible paths.
\begin{thm}
The proposed EEAS strategy achieves the
maximum diversity gain of $\alpha$ for both the $T_1 \rightarrow T_2$ and
$T_2 \rightarrow T_1$ communication, simultaneously in a two-way
multi-hop relay channel.
\end{thm}
\begin{proof}
We analyze the DM-tradeoff for $T_1$ to $T_2$ transmission (\ref{twowayrx})
using the proposed EEAS strategy.
Note that the DM-tradeoff analysis for $T_2$ to $T_1$ transmission
follows similarly, since, the path $p_1$ also
maximizes the product of the norm among all paths from $T_2$ to $T_1$ due to
the reciprocity assumption on the channel coefficients.

From Theorem $3.3.$ \cite{Sreeram2008},
it follows that the DM-tradeoff $d_{out, \bH}(r)$ with
the channel $\bH$ is lower bounded by the DM-tradeoff $d_{out, \bH_d}(r)$ with
channel $\bH_d$, where $\bH_d$ is a diagonal matrix that contains the diagonal
entries of $\bH$. Therefore, to obtain a lower bound on the DM-tradeoff
 of the two-way multi-hop relay channel with the proposed EEAS strategy,
we analyze its DM-tradeoff with channel $\bH_d$.
With channel $\bH_d$, (\ref{twowayrx}) can be rewritten as
\begin{eqnarray*}
\left[\begin{array}{c}
y^2_{1+N} \\ y^2_{2+N} \\ \vdots \\y^2_{T+N}
\end{array}\right]
 & = & \underbrace{\left[\begin{array}{cccccc}
\lambda_0 & 0 & 0 & 0 \\
0  & \lambda_0 & 0 &  0 \\
\vdots & \ddots & \ddots & \vdots \\
0 & 0 & 0&  \lambda_0
\end{array}\right]}_{\bH_d} \left[\begin{array}{c}
x^1_{1} \\ x^1_{2} \\ \vdots \\x^1_{T}
\end{array}\right]  +  \left[\begin{array}{c}
z^2_{1+N} \\ z^2_{2+N} \\ \vdots \\z^2_{T+N}
\end{array}\right],
\end{eqnarray*}
where $\lambda_0 = \prod_{n=0}^{N-1}\sqrt{\mu_nP}h^n_{i^*_ni^*_{n+1}}$.
It can be shown that the DM-tradeoff with channel $\bH_d$ is same as the DM-tradeoff of the scalar channel with input output relation
\[y^2_{t+N} = \prod_{n=0}^{N-1}\sqrt{\mu_nP}h^n_{i^*_ni^*_{n+1}}x^1_t + z^2_{t+N}, \ \forall \ t=1,\ldots,T.\]
Let the variance of $z^2_{t+N}$ be $\sigma^2$ and $\SNR \bydef \frac{P\prod_{n=0}^{N-1}\mu_n}{\sigma^2}$, then the outage probability of signal $x^1_t$ received at $T_2$ at time $t+N$ is
\begin{eqnarray*}
P_{out}\left(r\log\SNR\right) &=& P\left(\log\left(1+\SNR\prod_{n=0}^{N-1}|h^n_{i^*_ni^*_{n+1}}|^2\right) \le r\log\SNR\right)
\end{eqnarray*}
\begin{eqnarray*}
P_{out}\left(r\log\SNR\right)
 &\expeq& P\left(\log\left(\max_{i_n\in{0,1,\ldots, M_n},
\ n\in\{0,1,\ldots,N-1\}} \SNR\prod_{n=0}^{N-1}
|h^n_{i_ni_{n+1}}|^2\right) \le r\log\SNR\right).
\end{eqnarray*}
Clearly,
\begin{eqnarray*}
P_{out}(r\log\SNR) &\le &
P\left(\max_{\mathbb{P}_N} \SNR\prod_{n=0}^{N-1}
|h^n_{i_ni_{n+1}}|^2
\le
r\log\SNR\right),
\end{eqnarray*}
where $\mathbb{P}_N$ is the set containing maximum number of independent paths in a
multi-hop relay channel.
From Lemma \ref{maxindepath}, $|\mathbb{P}_N|$ is
$\alpha$ and by definition of independent paths,
 channel coefficients on independent paths are independent.
Thus,
\begin{eqnarray*}
P_{out}(r\log\SNR) &= & \prod_{j=1}^{\alpha}P\left(\SNR\prod_{n=0}^{N-1}
|h^n_{i_ni_{n+1}}|^2 \le
\SNR^{-(1-r)}\right). \\
\end{eqnarray*}
As described in Section \ref{sec:full-dup},
\[P\left(\prod_{n=0}^{N-1}|h^n_{i_ni_{n+1}}|^2 \le
\SNR^{-(1-r)}\right) \expeq
\SNR^{(1-r)}, \ r\le 1.\]
Thus
\[P_{out}(r\log\SNR) \expl  \SNR^{-\alpha(1-r)}, \ r\le 1\] and
\[d_{out}(r) = \alpha(1-r), r \le 1.\]
Similar expression can be obtained for $T_2$ to $T_1$ communication.
Since $d(r) = \alpha(1-r), \ r \le 1$, the maximum diversity
gain of EEAS strategy (obtained at $r=0$) is
$\alpha = \min_{n=0,1,\ldots,N-1}\{M_nM_{n+1}\}$ for $T_1\rightarrow T_2$ and
$T_2\rightarrow T_1$ communication, simultaneously,
which equals the upper bound on diversity gain from Lemma \ref{upbounddmt}.
\end{proof}

\begin{thm}
The proposed EEAS strategy with AF at each relay achieves the
maximum multiplexing gain in multi-hop relay channel for the
$T_1\rightarrow T_2$ and $T_2\rightarrow T_1$ communication, simultaneously,
if either the source,
or any of the relay stages or the destination has only a single antenna.
\end{thm}
\begin{proof} Since $d(r) = d_{out}(r) = \alpha(1-r), r \le 1$, the maximum
multiplexing achievable multiplexing gain is $1$ which equals the upper bound
on multiplexing gain from Lemma \ref{upbounddmt}.
\end{proof}

{\it Discussion:} Our proposed EEAS strategy achieves the maximum
diversity gain of $\alpha$, simultaneously for $T_1$ to $T_2$
and $T_2$ to $T_1$ communication.
The result is primarily due to the reciprocity assumption on the channel
coefficients between adjacent stages, because with this assumption, the
path chosen for $T_1\rightarrow T_2$ communication is also the "best"
path from $T_2\rightarrow T_1$. The reciprocity assumption is valid in a
time-division duplex system, when the coherence time is larger than the
number of relay stages $N$.

Thus, this result implies that by using the proposed EEAS strategy
both $T_1$ and $T_2$ can exchange messages,
with each other, simultaneously, without any loss in diversity gain or
multiplexing gain as compared to only one-way communication
(Section \ref{sec:full-dup}).

\section{Simulation Results}
\label{sec:sim}
In this section we provide some simulation
results to demonstrate the uncoded bit error rates (BER) of the EEAS strategy
and compare its performance with respect to the DSTBCs proposed in
\cite{Jing2004d} and \cite{Vaze2008}. The specific cases of $N=2$ and $N=3$
are considered.
To have a fair comparison with the DSTBCs \cite{Jing2004d} and \cite{Vaze2008},
 in all the simulation plots $P$ denotes the total power used by all nodes in
the multi-hop relay channel. In all the simulation plots we
use $4$ QAM modulation.

\begin{figure}[h]
\centering
\includegraphics[height= 2.6in]{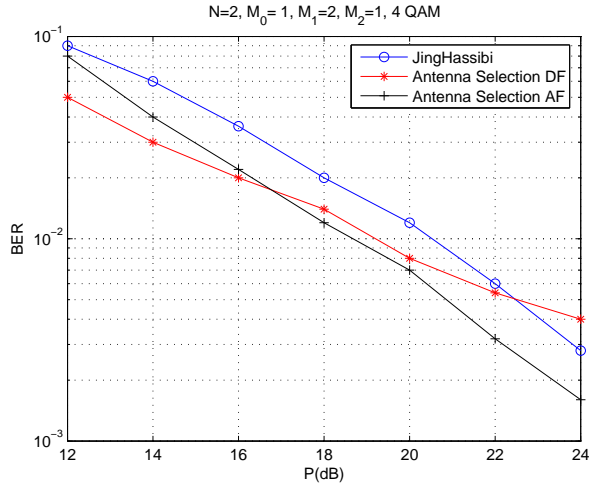}
\caption{BER comparison of EEAS strategy with JingHassibi
code for $N=2, M_0=1, M_1=2, M_2=1.$} \label{2stage}
\end{figure}

In Fig. \ref{2stage} we plot the BER of the EEAS and the
comparable DSTBC from \cite{Jing2004d} for $N=2$, $M_0=1, M_1=2$ and $M_2=1$.
For comparison purposes, we also plot BER for the EEAS with DF
at each antenna on the path selected by the EEAS strategy.
It is easy to see that the EEAS
and the DSTBC \cite{Jing2004d} achieve the maximum diversity gain.
Moreover, EEAS with AF requires $2$ dB less power than the
DSTBC to achieve the same BER.

\begin{figure}[h]
\centering
\includegraphics[height= 2.6in]{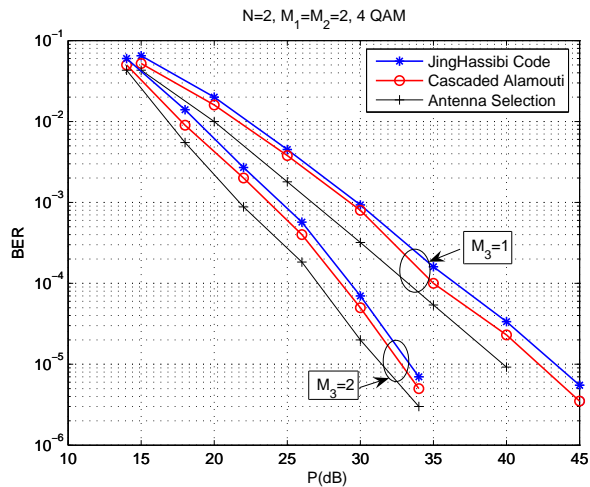}
\caption{BER comparison of EEAS strategy with JingHassibi and cascaded
Alamouti code for $N=2, M_0=M_1=2$.}
\label{powercomp}
\end{figure}

Next we plot the BER curves for $N=2$, $M_0=M_1=2$ with $M_2=1,2$ in Fig.
\ref{powercomp} for the EEAS strategy and the DSTBCs \cite{Jing2004d,Vaze2008}.
The DSTBC \cite{Vaze2008} is denoted as the
cascaded Alamouti code. For this configuration also, it is easy to see that
both the EEAS strategy and the DSTBCs \cite{Jing2004d,Vaze2008} achieve the
maximum diversity gain, however, EEAS requires $2$ dB less power than the
DSTBC to achieve the same BERs.

\begin{figure}
\centering
\includegraphics[height= 2.4in]{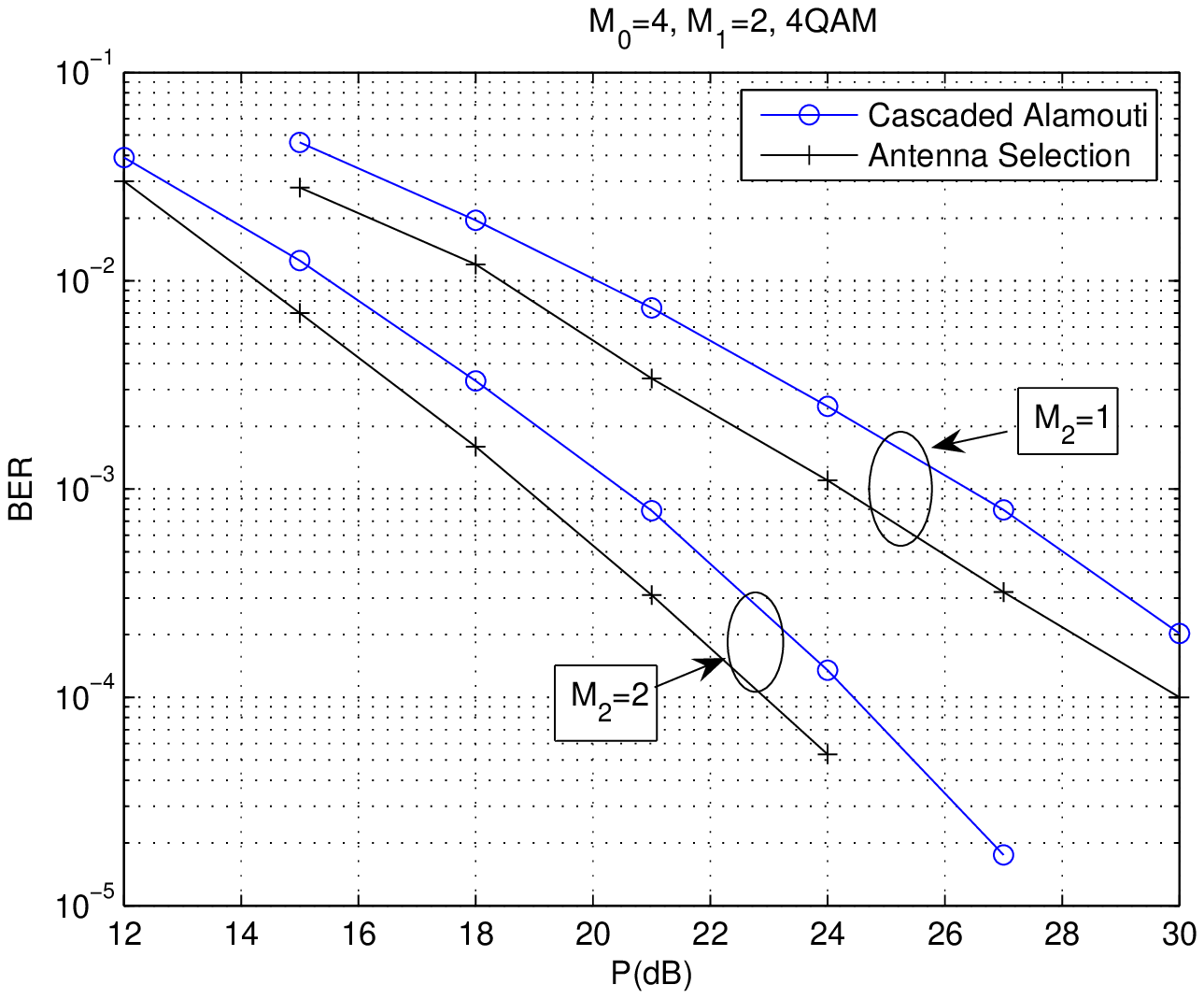}
\caption{BER comparison of EEAS strategy with cascaded
Alamouti code for $N=2, M_0=4, M_1=2$.}
\label{2stage4x2}
\end{figure}

Then we plot BER for the EEAS strategy for $N=2$, $M_0=4, \ M_1=2$ and $M_2=1,2$ in Fig.
\ref{2stage4x2}  and compare it with the DSTBC \cite{Vaze2008} which is denoted as cascaded OSTBC. From Fig. \ref{2stage4x2} it is clear that both the EEAS
strategy and the DSTBC achieve the maximum diversity gain, but EEAS requires
about $2$ dB less power to achieve the same BER.

\begin{figure}
\centering
\includegraphics[height= 2.4in]{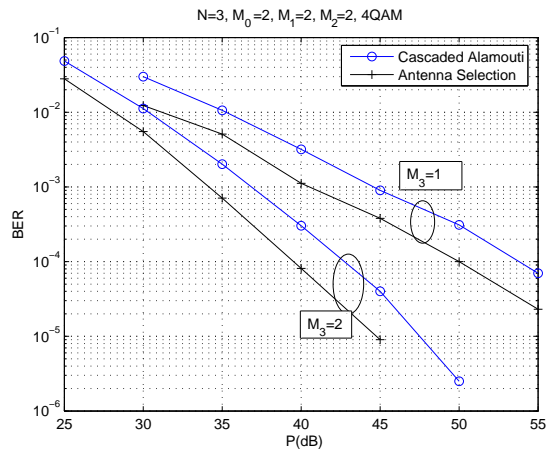}
\caption{BER comparison of EEAS strategy with cascaded
Alamouti code for $N=3, M_0=M_1=M_2=2$.}
\label{3stage}
\end{figure}

Finally, in Fig. \ref{3stage} we plot the BER of the EEAS strategy and
 the cascaded Alamouti code \cite{Vaze2008} for $N=3$ where $M_0=M_1=M_2=2$
with  $M_3=1,2$. In this case also,
it is clear that the EEAS and the cascaded Alamouti code achieve the maximum
diversity gain, but there is a $5$ dB SNR gain for EEAS strategy over the
cascaded Alamouti code \cite{Vaze2008}.

The improved BER performance of EEAS over DSTBCs \cite{Jing2004d, Vaze2008},
is due to fact that with EEAS strategy only one antenna from each relay stage
is used for transmission.
The advantages of using only one antenna with the EEAS strategy are two-fold.
Firstly, all the power dedicated to a relay stage is transmitted by a
single antenna rather than being equally divided among all the antennas of the
relay stage as is the case in DSTBCs, and thus, improves the signal power at
the destination.
Secondly, since only one antenna is used for forwarding the signal from the
source to the destination, the effective noise power at the destination is smaller
as compared to DSTBCs where each antenna of each relay is used for forwarding
the signal.


\section{Conclusions}
\label{sec:conc}
In this paper we considered the problem of maximizing
the diversity gain in a multi-hop relay channel. We proposed different EEAS
strategies for full-duplex one-way, half-duplex one-way and
full-duplex two-way multi-hop relay channels. For the case of full-duplex,
one-way and two-way multi-hop relay channels we showed that the proposed
EEAS strategies achieve the maximum diversity gain.
For the half-duplex one-way case, we showed that the proposed EEAS strategy
achieves the full-duplex rate of transmission
with a minimal loss in the diversity gain.

The main conclusion we derived in this paper is that with EEAS
strategies maximum diversity gain can be achieved in a multi-hop relay channel
without any space-time coding (DSTBC). Next, we present a brief comparison 
of both these strategies with respect to several important performance metrics. 
\begin{itemize}
\item Overhead: In the case of DSTBC's, CSI
is required at each relay node and at the destination in the receive mode.
 With EEAS, CSI is only needed at the destination in the receive mode.
In this case, however, a low bit-rate feedback is required from the destination to the source and each relay stage
 to communicate the source and the relay stage antenna indices to use.
 Thus EEAS reduces the training overhead compared to DSTBCs, but requires a low-rate feedback link.

\item Multiplexing gain \cite{Zheng2003}:
The full-rate DSTBC \cite{Yang2007a} achieves the maximum multiplexing gain for a
general multi-hop relay channel, while the DSTBC \cite{Sreeram2008,Vaze2008}
and the proposed EEAS achieves the maximum multiplexing gain for multi-hop relay
channels where either the source or any relay stage or the destination has a
single antenna.

\item Synchronization: For DSTBCs, perfect frame synchronization is required at
each relay node of each relay stage, which is a very strict requirement and
hard to meet in practice. With the EEAS strategies, however,
due to the use of only a single antenna at each relay stage frame synchronization is easy,
since different relay nodes of a relay stage need not be perfectly frame synchronized.

\item Network Resource Utilization: With DSTBCs, each antenna of every relay node is used to achieve the maximum
diversity gain. In contrast, with EEAS,
only $N+1$ antennas are used, one each from the source, each relay stage
and the destination.
Thus, with EEAS most of the relay nodes are unused and can either enter sleep mode (important for power limited nodes)
or serve as relays for other links in the network, subject to not interfering with the primary communication.

\item Noise Amplification: To achieve maximum diversity gain with DSTBC,
an AF strategy is used at each
relay node \cite{Yang2007a, Sreeram2008}. Therefore, with DSTBCs,
the noise received by all the relays in a relay stage is amplified
and forwarded to the next relay stage. With a large number of relay stages,
the contribution of the forwarded noise is significant in the received signal
at the destination and severely limits the SNR.
Using EEAS, only noise
received by a single antenna of each relay stage is forwarded to the next
relay stage and results in relatively less noise power at destination and
provides with a substantial array gain.
\item Decoding Complexity: Due to coding in space and time, the decoding
complexity of DSTBC is significant, except for the DSTBC \cite{Vaze2008} where the decoding complexity is minimal.
With the EEAS strategy, however, the decoding complexity is minimized,
since only one symbol is received by the destination at any given time instant.
\item Latency: With DSTBCs, coding is done in space as well as time.
Therefore, to decode the signal, the destination
has to wait for the full coding length before it can start decoding.
With multiple relay stages this delay is significant
and is not preferable for low-latency application such as voice communication.
With the EEAS strategy, however, the destination can decode the signal after
$N$ time slots, which is minimum possible, since the destination cannot
be reached from the source in less than $N$ time slots.
\end{itemize}

Thus, it is clear that the proposed EEAS strategies achieve the 
maximum diversity gain
in a multi-hop relay channel and provide several advantages over DSTBCs, 
however, they fail to achieve the maximum multiplexing gain. 
Since high data rates are required in current wireless systems, it will be 
of interest to see whether one can design EEAS strategies which also 
achieve the maximum multiplexing gain. 
Another metric of interest in multi-hop relay
channels or large wireless networks is the transmission capacity which is
defined as the number of source destination links in a network that can be
supported simultaneously. An interesting question to ask, which is beyond the
scope of the present paper is, how does the EEAS strategies affect the 
transmission capacity in a wireless network?

\bibliographystyle{IEEEtran}
\bibliography{IEEEabrv,Research}

\end{document}

%% file: input.tex
\usepackage{amsfonts}
\usepackage{times}
\usepackage{latexsym}
\usepackage{amssymb}
\usepackage{amsmath}
\usepackage{cite}
\usepackage{verbatim}

\newcommand{\bydef}{\triangleq}
\newcommand{\tr}{{\it{tr}}}
\def\SNR{{\textsf{SNR}}}

\def\bydef{:=}

\def\bb0{{\mathbb{0}}}

\def\bydef{:=}
\def\ba{{\mathbf{a}}}
\def\bb{{\mathbf{b}}}

\def\bh{{\mathbf{h}}}

\def\bn{{\mathbf{n}}}

\def\br{{\mathbf{r}}}
\def\bs{{\mathbf{s}}}

\def\bv{{\mathbf{v}}}

\def\bx{{\mathbf{x}}}

\def\b0{{\mathbf{0}}}

\def\bA{{\mathbf{A}}}

\def\bH{{\mathbf{H}}}

\def\bbC{{\mathbb{C}}}

\def\bbE{{\mathbb{E}}}

\def\bbN{{\mathbb{N}}}

\def\bbR{{\mathbb{R}}}

\def\bydef{:=}

\def\sf0{{\mathsf{0}}}

\def\Nt{{N_t}}
\def\Nr{{N_r}}
